# Temperature Profile Measurements During Heat Treatment of BSCCO 2212 Coils

Alvin Tollestrup
VHFSMC,
FNAL April 14, 2011

## Abstract

The temperature profile of two different BSCCO 2212 (1, 2) coils has been analyzed. The profiles are obtained from thermocouples imbedded in the windings during the heat treatment that activates the 2212. The melting and freezing of the 2212 is clearly observed. A model that describes the data and can be used to guide the processing of new coils has been developed.

## 1.0 Overview

We have obtained the thermal history of two BSCCO coils, one from NHMFL (1) that had 10 layers of 1 mm diameter wire with 0.15 mm insulation and a second coil from OST that had 24 layers with similar insulation and conductor size. Both coils had thermocouples imbedded in the windings and excellent recordings of the temperature over the whole reaction cycle were available for analysis.

There are several features that we will address in this note. Measurements have shown that the $I_c$ of the conductor is a sensitive function of its thermal history. This brings up the question of the absolute accuracy of the thermometry in the range around 882 $^0$C, the MP of 2212. The reference for the treatment profile is really related to this MP and to small deviations around it. Since the heat of fusion of 2212 is rather large, it generates a clear signal during the melting and cooling transition that automatically generates the relative temperature markers. The physics is the same as the way ice in water maintains an isothermal environment until it is all melted.

A related question is the thermal response time of the coil package. The temperature cycles that are being used to optimize strand and small coils can have rapid changes easily implemented whereas a large coil may have such a large thermal time constant that the optimum cycle may not be attainable. A simple analytical model that works well for small solenoids has been developed and an ANSYS (5) program that works for larger coils with more complicated geometry has been set up but will not be discussed in this note.

## 2.0 A Toy to Set the Stage

Before analyzing some real data, it is useful to look at a very simple case that is easy to solve. We consider a slab of material that is 1 unit thick and with a heat source that linearly increases the temperature of the two faces. This is shown in Fig.1.

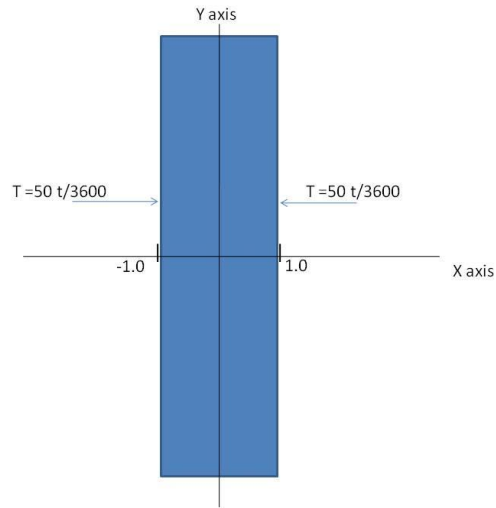

Fig.2.1   Toy slab with the sides heated at 50 $^{\circ}$C /hour.

The slab has the thermal constants about the same as pure silver:  Cp = 300 J/kg, heat conductivity k = 450 W m$^{-1}$ K$^{-1}$, and density = 10.49 g cm$^{3}$ . There is one additional feature we need to model and that is the heat of fusion when the material melts. To simulate this, we apply an approximation of a delta function to Cp at 50 $^{\circ}$C. The area under the delta function is equal to the heat of fusion. Delta functions make difficulties in solving the heat equation and so we smooth it out to have a Gaussian shape with small width.

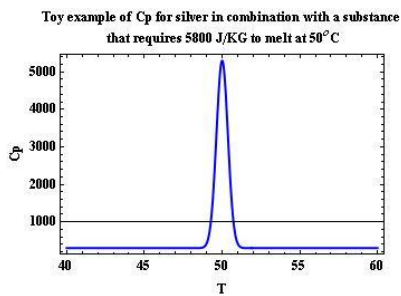

Fig 2.2  Cp of a material that melts at 50 $^{\circ}$C

The distribution of temperature in the slab has been solved in Mathematica and the results are shown in Fig.2.3 below.

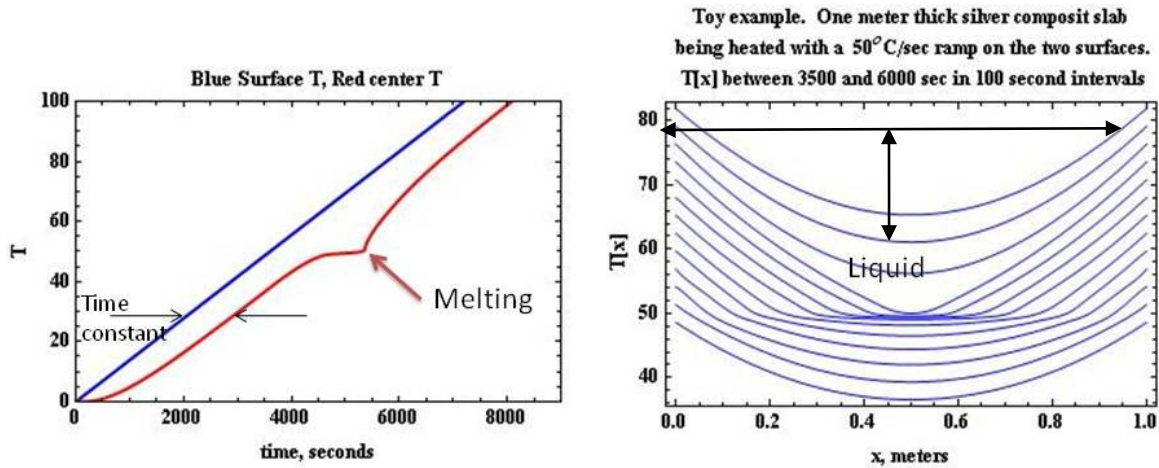

Fig.2.3 Left: Temperature variation with time. The blue curve is T[t] of the two faces of the slab, the red curve is the middle. The time shift between the two curves is the time constant of the material. On the right is the temperature profile thru the slab showing the melting progressing from the outside to the center. The material is all melted for the top three curves.

Several interesting things can be learned from this toy example. First, it is important to understand and have a simple measurement of the transient response of the coil mass to ensure that it responds to the desired temperature profile. If the requested profile has regions that change faster than the response time of the coil, the BSCCO will not achieve its optimum Ic. A uniform ramp of the oven temperature plus a thermocouple placed in the center of the coil allows a direct in situ measurement in real time of the response time constant as shown in the left side of Fig.2.3 The diffusion equation is shown below, where k is the conductivity, Cp the heat capacity, and ρ is the density.

$$\partial_t T[x, t] = \frac{k}{C_p \rho} \nabla^2 T[x, t]$$

$$T[x,t] = \alpha\, t - \beta \left(1 - \left(\frac{x}{a}\right)^2\right)$$

$$\beta = \alpha\, \frac{\rho\, C_p}{2k}\, a^2$$

$$\tau = \frac{\rho\, C_p}{2k}\, a^2$$

The particular solution for T[x,t] shown on the second line is when things come into equilibrium on a linear ramp with rate of rise $\alpha$. $\beta$ is the depth of the parabolic dip in temperature at the slab center relative to the faces being heated. It can be derived by inserting the particular solution into the heat equation.

The time constant $\tau$ is given by $\beta/\alpha$. The time constant gives us a measure of the length of time the coil package takes to come into equilibrium after a change in the heat program. This exponential change is seen near t=0 at the start of the ramp in the left hand side of Fig.2.3.

It is important to note the usefulness of information derived from the linear ramp. Each layer of the coil is increasing its temperature linearly. Since we can calculate the heat capacity of the layer from the known Cp of the constituents, we know exactly how many joules / second are crossing between layers. Since this is the same for each layer, the change in gradient across each insulation layer is known which leads to the parabolic heat distribution. An additional point worth noting is that adding a term linear in x to T[x,t] does not change things as the 2$^{nd}$ order derivative in x eliminates this term. Such a linear term allows for a heat difference between the two sides of the coil or in the case of a solenoid between the inside and outside layer of the coil. This will become apparent in a following example. However, the measure we have used above using the time delay between the outside and center of the winding is a good overall estimate of how fast the coil package can respond to temperature changes. This has been verified by analytically solving the heat equation in cylindrical coordinates and comparing with the slab solution above.

These results have been illustrated using a slab, but they are true in general. For a real coil, the heat capacity is averaged over the packing fraction of the conductor. The heat conductivity k is controlled by the insulation at low temperatures and by radiative transfer at temperatures around 800 $^0$C and above. More insulation will result in longer response times. We have developed an ANSYS [5] program that accurately models the coil winding structure, but the simple physics outlined above with the help of a discrete model using difference equations is probably sufficient to cover most problems at present.

> *What have we learned from this toy example?*
> 1. *The delay on a linear ramp between the surface and center of a coil gives us a good approximation to the coil response time.*
> 2. *The $\tau$ becomes larger like the square of the thickness or number of layers.*
> 3. *$\tau$ can be easily measured in situ on the initial fast up ramp. It can be used to predict the temperature difference on the very slow ramps used in the 2212 formation during cooling, ie delta T = $\alpha\,\tau$ since $\alpha$ is accurately known from the furnace program.*

## 3.0 Transient Response of the NHMFL 10 Layer Coil Package.

We will first look at the transient response of the coil package as measured by the TCs. The temperature profile used for oven control for the coil from NHMFL coil is shown in Fig. 3.1 bellow. It offers a number of opportunities to measure the properties of the system.

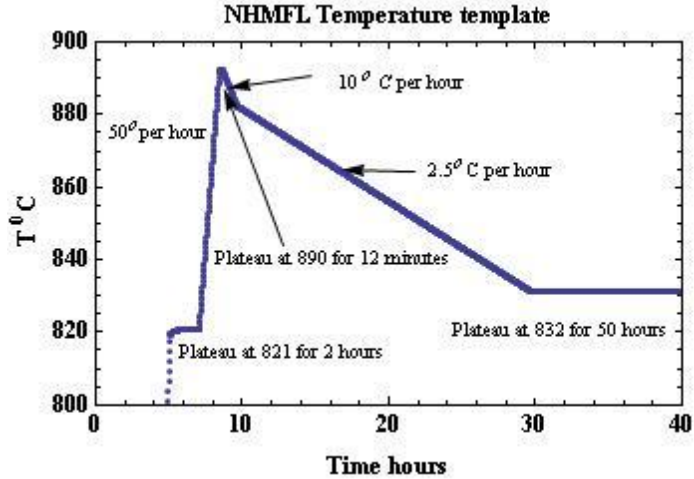

Fig. 3.1 Temperature profile of the oven during heat treatment of the NHMFL coil

The first thing that has to be done is to intercalibrate the TCs and find their RMS fluctuations. The long plateau after 30 hours offers an ideal place to intercalibrate as well as to study the standard deviation of the temperature readings. The results are shown in the following Table.

| TC # | #1 | #2 | #3 | #4 | #5 | #6 | #7 | #8 |
|---|---|---|---|---|---|---|---|---|
| Radius mm | 0.0 | 14.31 | 16.81 | 19.81 | 23.31 | 23.81 | 25.81 | 28.7 |
| Mean T $^0$C | 831.26 | 832.30 | 832.46 | 832.45 | 832.50 | 832.35 | 832.48 | 832.74 |
| Delta T $^0$C | -1.2 | -.17 | -.006 | -.02 | -.03 | -.12 | .009 | .28 |
| STD $^0$C | 0.043 | 0.039 | 0.057 | 0.056 | 0.056 | 0.062 | 0.035 | 0.029 |

The TC#1 is on axis and seems to respond most closely to the oven program although it is systematically low by 1.2$^0$C. As a result, we have taken all of the other TCs during the time period between 30 and 56 hours and formed an average of their readings which is 832.467 $^0$C and using this average we have generated the delta T's shown above. The last row in the table shows the Standard Deviation of the individual TC's. This shows that the recording system is really capable of excellent resolution and it is worth recording at least two decimal places when curing a coil. We also note that the absolute calibration of the individual TCs is really quite good as can be seen in the Mean T before correction. (Note: This assumes there was no correction applied before this run!).

An interesting question to ask is does the normalization of the TCs on the rear plateau apply to the front plateau. The Fig.5 below shows the comparison. The left hand figure shows that there is a systematic drift of the oven temperature (assumed to be given by TC#1) upward by about $0.6^0C$ during about 1.5 hours (The first 0.5 hours of this plateau has been eliminated due to the transient response of the coil). The plot on the right shows the difference between four of the TCs as "calibrated on the rear plateau" and the furnace, a number which should be zero except for noise fluctuations. It appears that the calibration is stable to less than $0.2^0C$ during a period of about 60 hours.

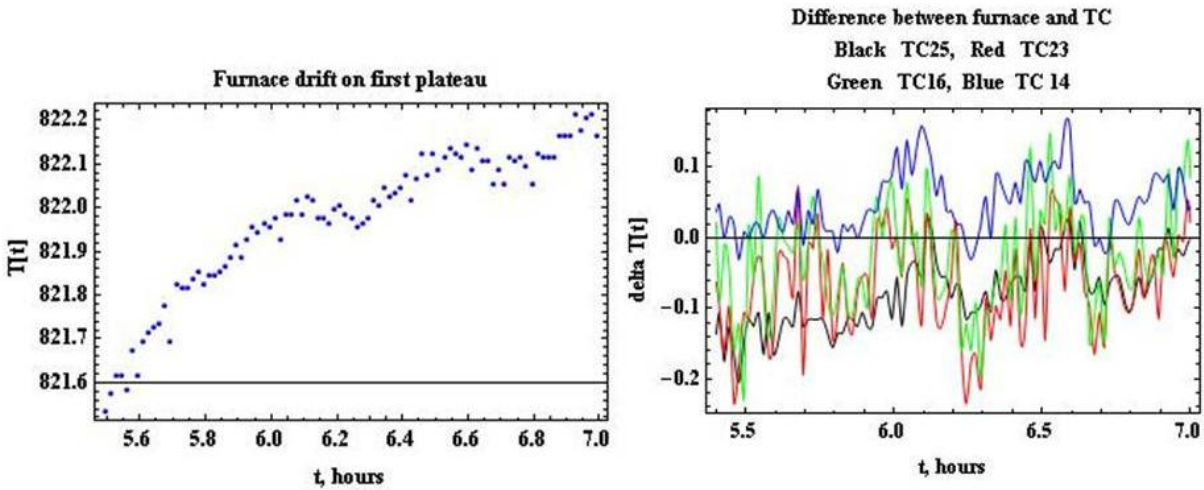

Fig.3.2. The left hand plot shows the TC at r=0 drifting during the front plateau. The right plot shows the difference between four of the other TCs and the one at r=0 during the same period.

The calibration system that we have used assumes that the desired differences of oven temperature in the region $800-900^0C$ are properly generated by the oven and mirrored in the response of the TC at r=0. We then bring the whole ensemble of TCs into agreement with the normalization process used above. This leaves in question the absolute calibration, but as we show below this is unimportant as there is a good internal indication of the melting point, MP, of BSCCO 2212 and this is the only important temperature on which the curing cycle should be keyed.

## 3.1 Transient response of the coil package.

There are several places where the transient response of the coil package can be studied. Referring back to Section 2 we can use the linear ramp section of the curing cycle to get a measurement of the package time constant. We can also use the way the temperature comes into equilibrium at the junction of a linear ramp and a plateau to give a measurement of this number.

The time = 77.6 hours is when the final cool down starts and provides a rapid down ramp and a simple place to test for time shifts. Fig. 3.3 on the next page shows the behavior at the start of the change.  We know from Section 2 that the center of the coil package should lag the oven temperature by a coil time constant.  The blue curve is the oven temperature, the red curve is the oven temperature shifted by 0.048 hours and the yellow curve is the response of the TC at 19.81 mm and yields a time constant of slightly less than 3 minutes.  The difference between red and yellow is due to the transient exponential response of the coil package before it comes into equilibrium with the falling oven temperature.

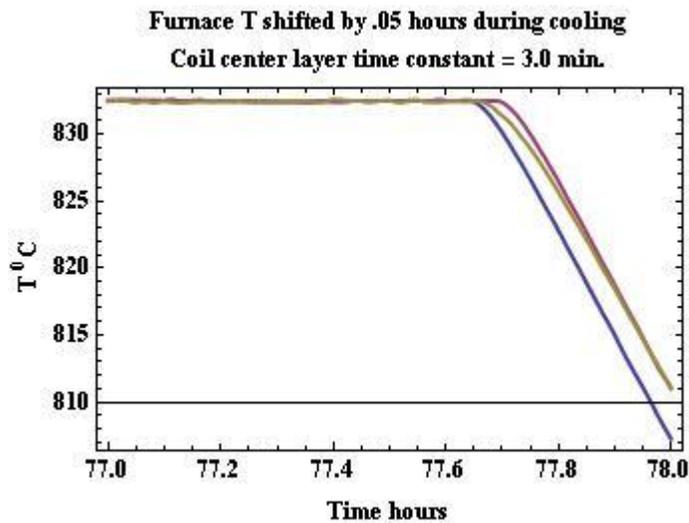

Fig. 3.3.  Blue….Oven Temperature, Red…….Oven shifted by .048 hours, Yellow….Coil response TC 19.81 mm.

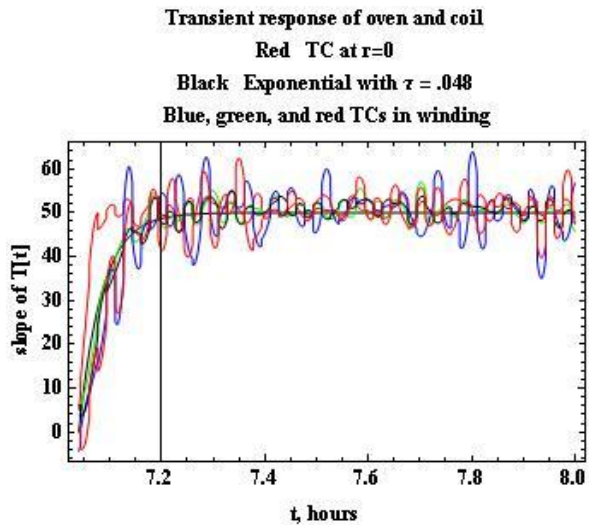

Fig. 3.4.  Transient response to a step change in the slope of the oven temperature.  The Red curve is the slope of the oven temperature.  The black is an exponential with a time constant of 0.048 hours.  The other curves are TCs in the coil package.

As mentioned above, a second place to look for the transient behavior is to examine the response when the dT/dt changes from zero to 50 $^0$C per hour, just after the front plateau at 821 $^0$C  The oven change is a step function in dT/dt and we would expect an exponential change in the slope of the coil package from zero on the plateau to 50 $^0$C per hour. Fig. 7, above, shows the response at 7 hours when a new ramp starts at dT/dt= 50 per hour. The black curve is a simple exponential with a time constant of .048 hours. The time constant from the time shift method is also a reasonable fit here. Remember that our goal is to get an approximate feeling for the coil thermal response in order to give us guidance that the desired temperature profile can be actually obtained in a real coil. Next, we look at the transient behavior of a coil during the region where the BSCCO melts.

## 3.2  The effect of BSCCO melting.

Using the calibration described above, we show in Fig.3.5 the temperature response of the 8 TCs on the rising part of the cycle where the 2212 melts. The vertical line indicates where the oven's peak temperature is reached and where it resides for 12 minutes before decreasing. The melt phase of the 2212 is clearly observed and is a bigger effect in the center than at the outer edges of the coil package. This effect comes from the absorption of heat during the melting process as explained in Toy Example in Section 2. The effect will become more pronounced in a linear way as the average thermal conductivity (primarily due to the electrical insulation) of the coil decreases and in a quadratic way as the number of layers increase. A simple model for the behavior will be developed below.

The top left plot is at r=0 and responds primarily to the oven temperature. The top and bottom right hand figures are from measurements on the inner and outer surfaces of the solenoid and the others are inside the coil and have their radius indicated at the top of the figure. The slope of the temperature curve during the melting process is probably caused by the fact that the TCs are not directly immersed in BSCCO but are somewhat isolated by the electrical insulation. The vertical line is at the time that the oven reaches its peak temperature of 890 $^0$C, where it remains for 12 minutes. It is interesting to note that 12 minutes was just barely enough for the whole coil to reach the peak temperature. Since the time constant scales as the number of layers squared, a coil with more layers would have not had all the layers reach the peak temperature.

It is clear that the melting of the BSCCO is clearly visible from the plots. A suggestion from this observation is that the temperature scale for the heat treatment cycle be set from this in situ measurement during the processing of the coil rather than relying on the absolute temperature calibration of the TCs. This is especially important as the peak temperature and the length of its duration are important parameters for maximizing the Jc of the conductor.

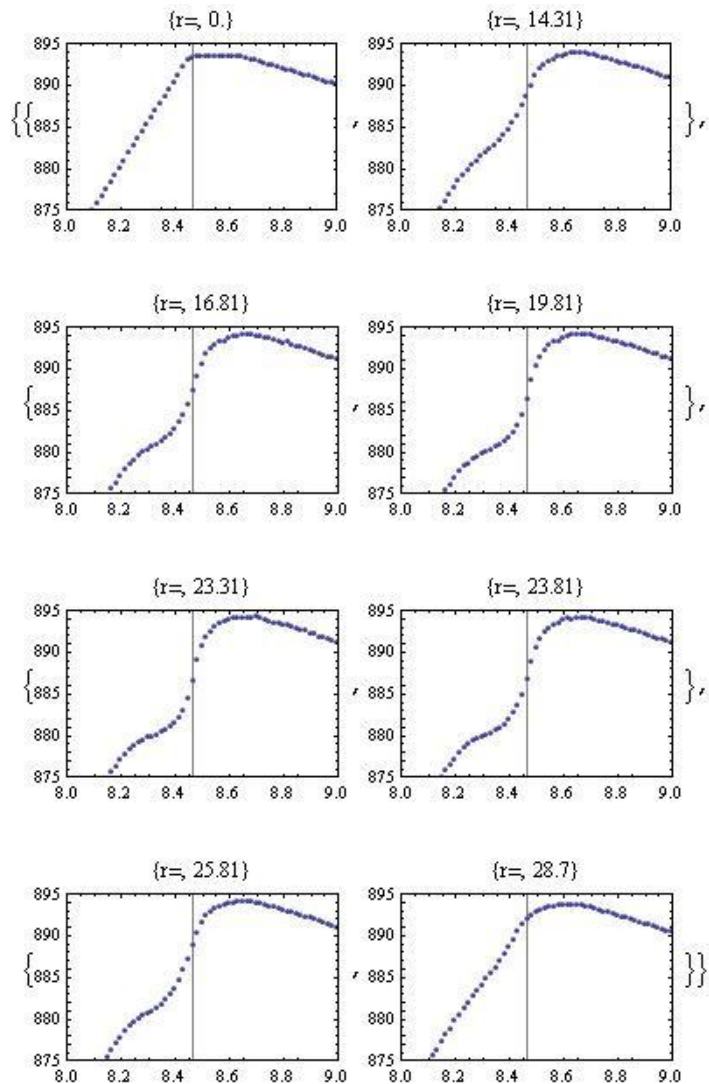

Fig.3.5. Temperature response of the 8 TCs. The radius in mm is shown above each plot. The one at r=0 is taken as the furnace temperature and the one at r=28.7 is taken to be on the outside of the coil

### 3.3 The freeze out of the BSCCO.

Having observed the melting of the BSCCO, the next useful observation would be if we could identify the freeze out. This measurement perhaps could actually help in optimizing the heat cycle. This is a more delicate measurement as the temperature gradient changes from 50 $^0$C per hour to 2.5 $^0$C per hour which will decrease the magnitude of the effect by a factor of 20.

Going back of Fig. 3.1 it is seen that there is a rapid drop in temperature at 10 $^0$C per hour from 892 $^0$C to 882 $^0$C at which point the rate of decrease changes to 2.5 $^0$C per hour. The freezing occurs at some point along this slow ramp which lasts for more than 20 hours. We could analyze this region by taking the difference between the oven TC and those inside the coil, but has the disadvantage that there is the statistical noise from two TCs. There is a long section of the ramp where the BSCCO has frozen and hence is no longer an energy source. A least square fit to the data from TCs inside the coil for the region between 18 and 28 hours (see Fig.3.1) can be made and extrapolated over the whole ramp. The region where freezing occurs will then show as a difference between this fit and the real data. Fig. 3.6 below shows the results.

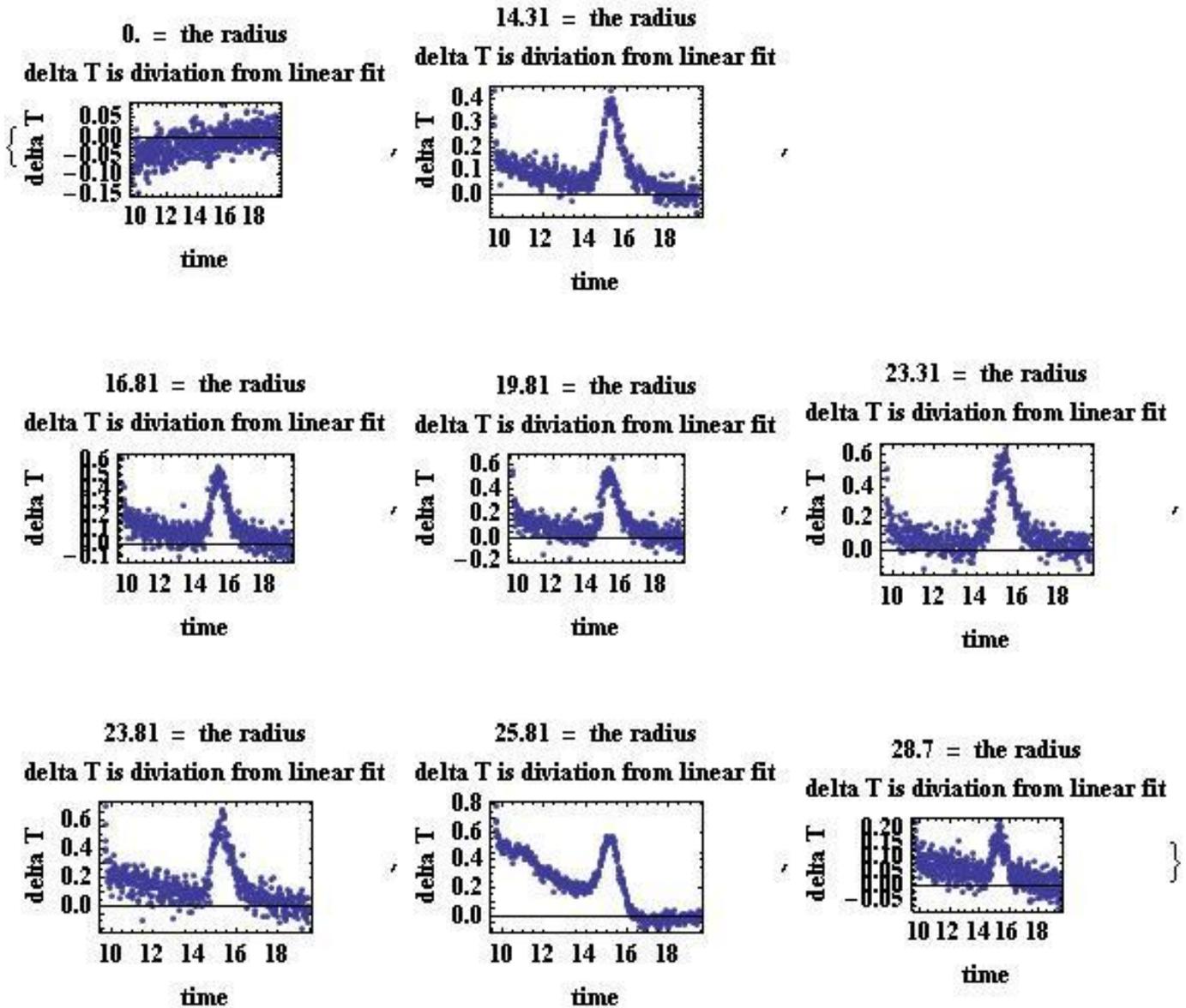

Fig. 3.6. These plots show the difference between an extrapolation of the temperature ramp generated from data between 18 and 28 hours and the real data. The radius of the TC in the coil is shown above each figure.

The freezing of the BSCCO keeps the local temperature above the ramp for a period of about one hour. We can replot the data above as deltaT vs T using the extrapolated ramp temperature and the result is shown in Fig. 3.7 on the next page where it is seen that the freeze out occurs at a temperature of 872 $^0$C or almost 10 $^0$C below the MP from Fig. 3.5

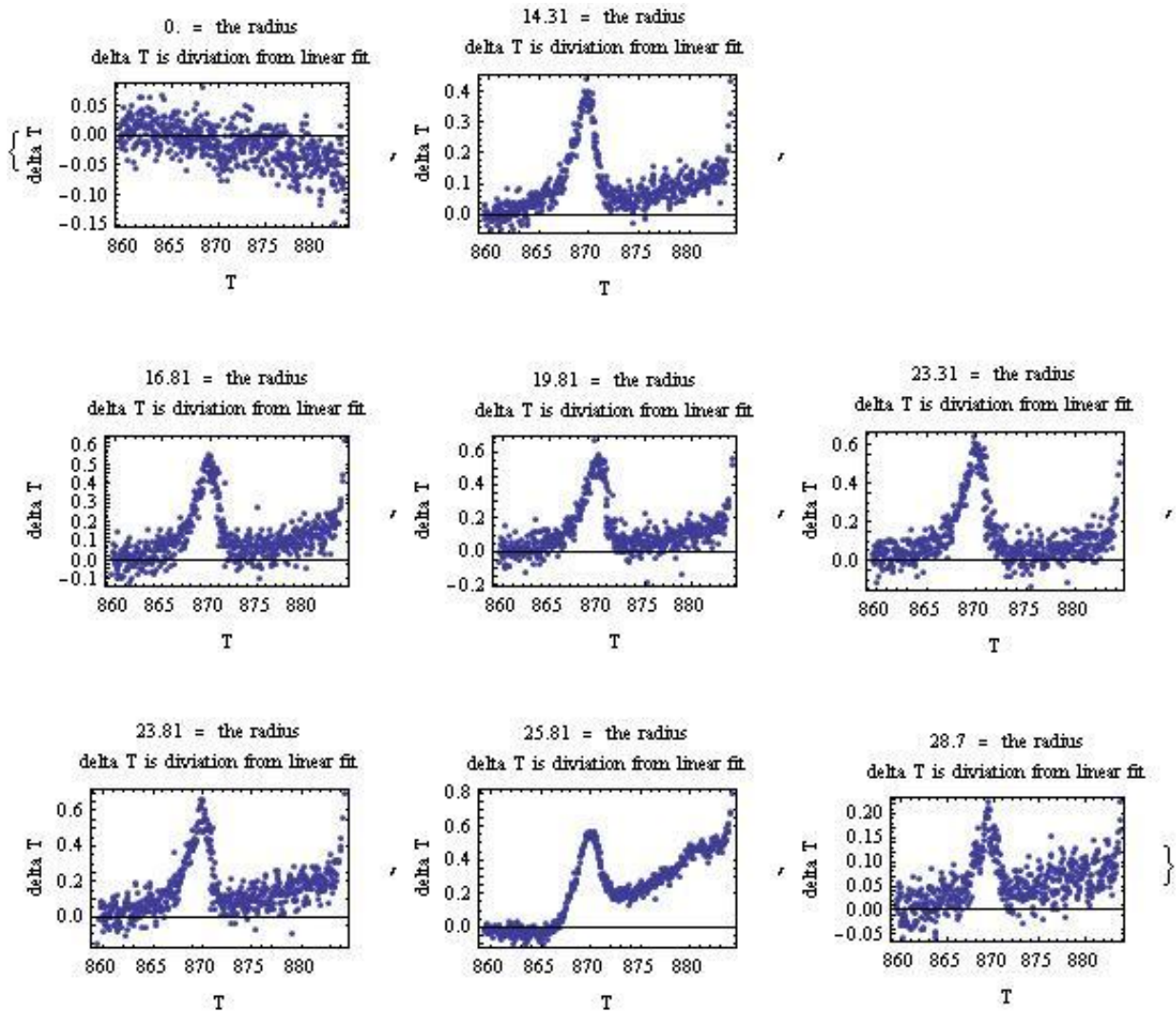

Fig. 3.7. This is the data of Fig. 9 plotted against temperature instead of time. (Time is increasing from right to left.) It is seen that the freezing takes place at a lower temperature than the melting. (See Fig. 8, where the MP is around 882 $^0$C. )

There is a little more information that we can extract. We can look at the temperature variation across the coil during the freezing at different times. We pick 14:55, 15:00, and 15:05 hours and show the temperature recorded by the eight TC across the coil at these three times in Fig. 3.8, below. The parabolic shape is seen, as expected

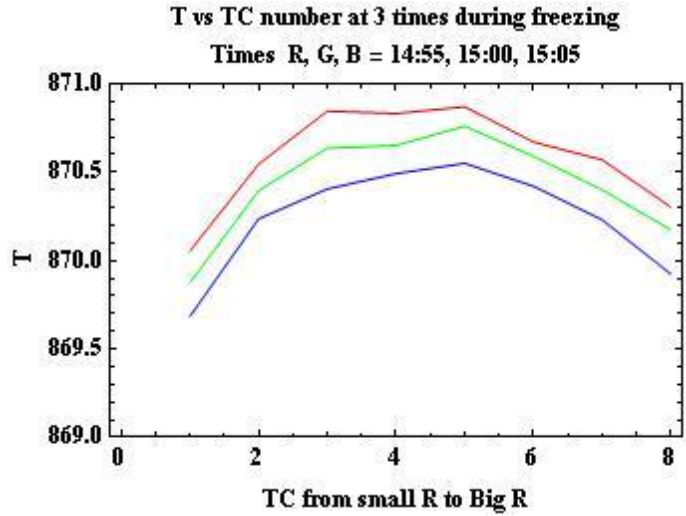

Fig. 3.8. Temperature recorded by the eight TCs across the coil at three different times decreasing from the peak.

## 4.0 Analysis of the 24 layer OST coil

OST very kindly supplied the thermal data obtained during the heat treatment of a 24 layer coil made with 1 mm strand and construction somewhat similar to the NHMFL coil just analyzed. With 24 layers, we might expect the coil time constant to be $2.4^2$ times longer. Measurements below show that its time constant is about 9-10 minutes compared to 3 minutes for the NHML coil. Apparently the conduction between layers is higher or it may be that these coils are so short that heating from the ends of the solenoid is also coming into play. In any case the longer time constant plus some details of the temperature cycle have made it more difficult to analyze. Never the less, the methods outlined yield useful results.

### 4.1 Normalization of the thermocouples.

This coil had 4 TCs, two were embedded in layers 9 and 15, one was placed in the center bore of the coil and the other on the outside of the package. The outer one seems to be the best monitor of the oven temperature as the one is the bore is influenced by the coil mass. For this analysis, a normalization of the three inner TCs has been made so that they all read the same as the outer one during the back plateau where the temperature is essentially held constant for a long period of time. This correction was {-1.19, -2.02, -.92} for TC {9, 15, coil axis}. This correction could be checked at other plateaus in the cycle and was stable to +/- $0.5^O$ C for the individual TCS, 9 and 15, but +/- $0.1^O$C for their average.

## 4.2 Coil transient response.

The thermal cycle of the coil for temperatures above 800° C is shown in the plot below.

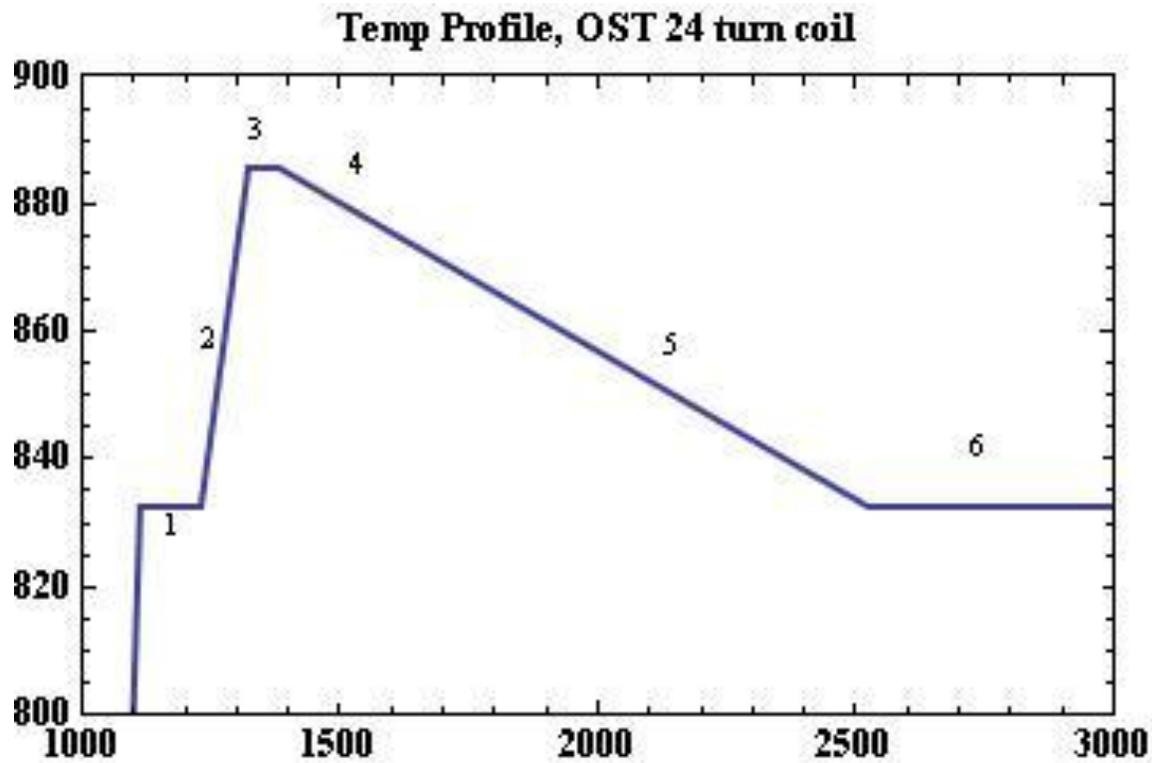

Fig. 4.1  Heat treatment temperature profile for the 24 layer OST coil. The various regions are numbered for future reference.

The best places to check for the transient response of the coil are where abrubt changes are made in the dT/dt and the region after the change responds exponentially with the local time constant.  A second measure is the temperature difference between the furnace and that inside the coil on a linear ramp where

$$\Delta T = \tau \, dT/dt.$$

A fit using both of these relations is shown for the start of the plateau in region 1 of Fig. 4.1. The time constant $\tau$ fixes both the amplitude of the exponential decay as well as its shape and the fit with $\tau$ = 10 minutes is quite good.

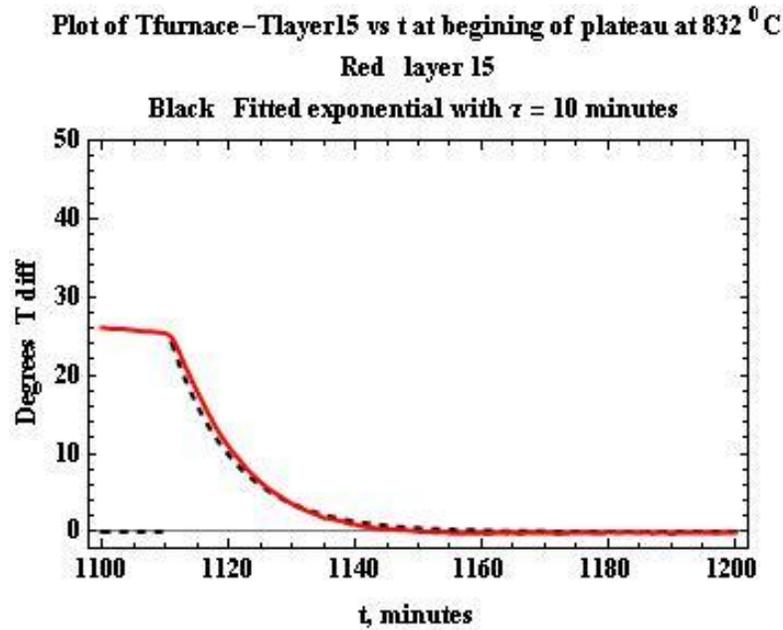

Fig. 4.2. The red curve is the temperature just before and for the next 80 minutes after the start of the plateau in region 1 of Fig. 4.1. The exponential starts when the furnace dT/dt = 0.

A second place to look is at the very end of the temperature when the furnace is turned off and the temperature starts to drop. This region is shown below. The red curve is the furnace temperature shifted forward by 9 minutes.

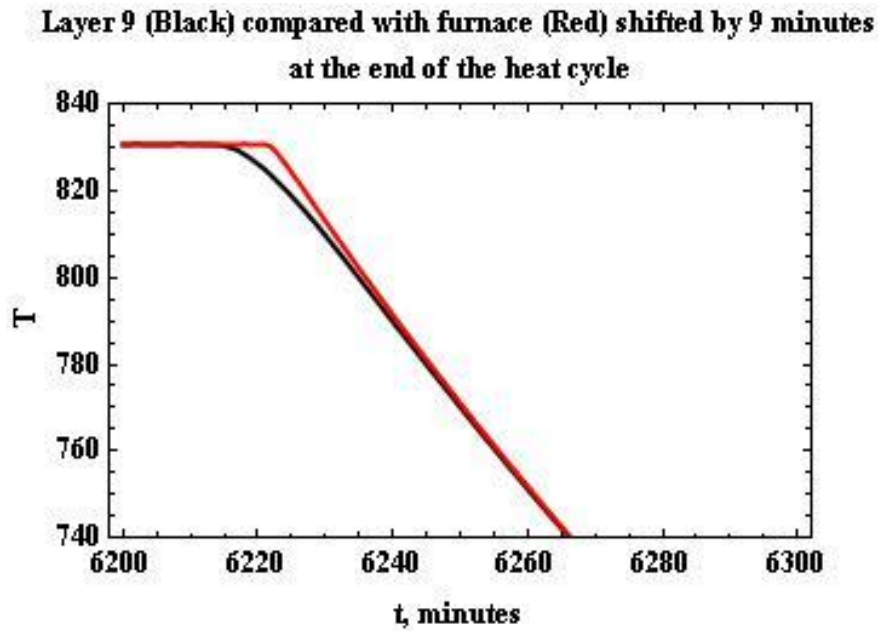

Fig.4.3. The red curve is the furnace temperature at the end of the profile but shifforward by 9 minutes and compared with the temperature of layer 9. Note that for falling furnace temperatures that the inside temperature lags in time and hence is hotter than the furnace.

The first part of the decline can be considered as a linear ramp and the coil time constant can be derived from the shift necessary to bring the two linear parts into coincidence. The answer of nine minutes compares pretty well with the number calculated for the fit shown in Fig. 4.2. A third place that we could check is at the start of the linear ramp from the plateau in region 1, Fig. 4.1. This worked well for the NHMFL coil, but we will shortly see a difficulty with using it here..

We can ask how does the τ for the OST coil compare with that measured at NHMFL? If the coils scaled with the number of layers squared as suggested in Sec. 2, we would expect 3 $(24/10)^2$ or about 18 minutes for the OST coil. Clearly, the OST coil has the better heat conduction into its interior. This may be due to different insulation between layers or a different aspect ratio due to the greater number of layers making conduction thru the ends more effective at heating the interior. For the present, we only need to look at the result as giving us a good measure of the transient response and as a guide for planning temperature profiles for future coils.

## 4.3 The effect of BSCCO melting.

Next, we consider region 2, the front linear ramp and its transition onto the plateau at the top temperature. In this region, the melting of the BSCCO clearly displayed itself as shown in Fig. 3.6 for the NHMFL coil. Fig. 4.4 below is a plot of the furnace temperature, layer 15, and a predicted curve using the measured time constants of the coil but assuming that the BSCCO doesn't melt.

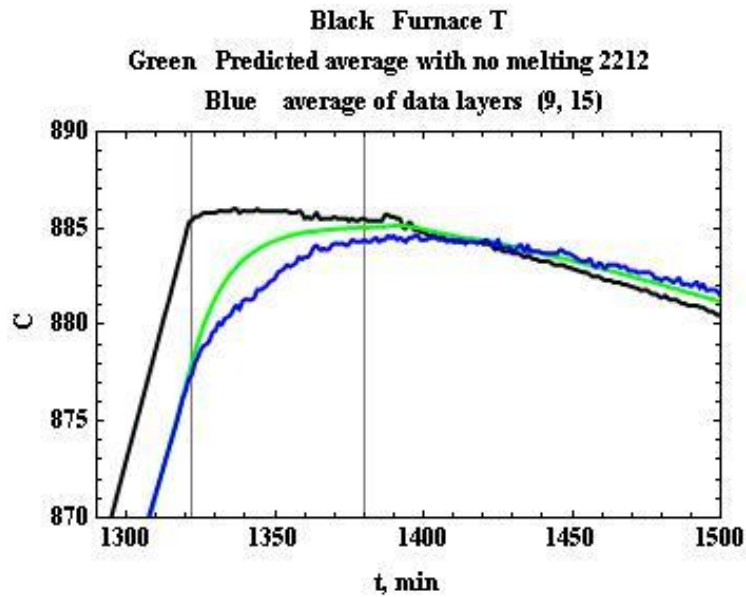

Fig. 4.4. The black curve is the oven temperature arriving at a plateau at t= 1322 minutes and then starting a slow decrease at 1380. The blue curve is the average temperature of layers 9 and 15 and should be compasred with the green curve which comes from a coil model where the BSCCO does not melt.

As the BSCCO starts to melt in layers 9 and 15, their temperature is clearly held down and the coil doesn't come into equilibrium until t= 1410 minutes.  The behavior is quite different than that displayed in the NHMFL coil mainly due to different peak temperatures in the two cases plus the longer time constant for the larger coil.

### 4.4  The freeze out of the BSCCO.

Fig. 4.5 shows the difference in temperature between the furnace and the average inside the coil over a wider range.

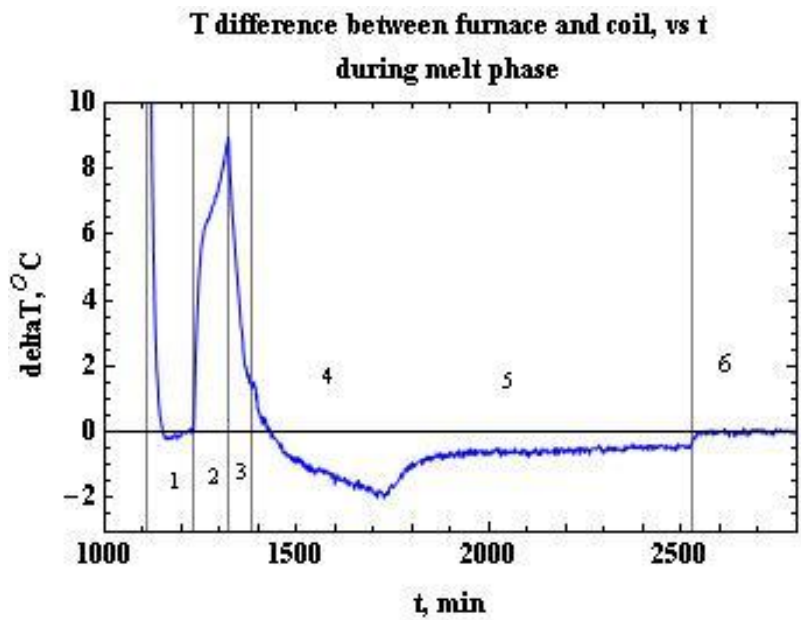

Fig. 4.5.  Furnace-coil temperature difference.  The number refer to the regions in Fig. 4.1

We have just discussed the regions 2, and 3 where the BSCCO melts.  Region 4 and 5 are on a cooling ramp of about -3.5 $^{o}$C / hour and the freeze out is clearly seen where the inside is maintained at an elevated temperature relative to that of the furnace.  The small difference after about 1750 is due to the the linear ramp and is consistant with the coil time constant and the ramp rate.  Another way to show this data is to plot things against temperature and is shown in Fig. 4.6.   The temperature is taken from the furnace profile and used as a parameter to map out the difference in T between the inside of the coil and the furnace.  Time flows to the right on the lower brance of the curve where the temperature is ramping up and the inside temperature is laging that of the furnace.  The $T_{max}$ plateau allows time for all of the 2212 to melt, followed by the freesing on the downward ramp which keeps the inside hotter than the furnace.

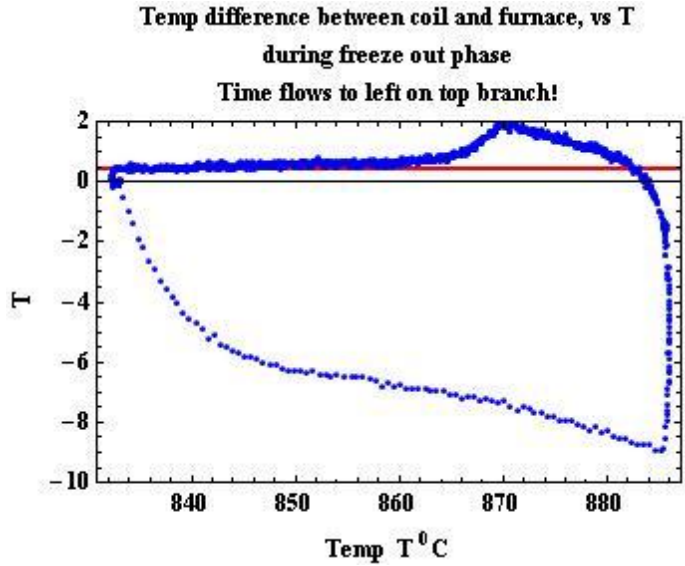

Fig. 4.6. Temperature difference coil – furnace vs furnace temperature. Time flows to the right on the lower branch where the 2212 is melting and to the left on the top where it is freezing.

## 5. Simple Thermal Model

WE now develop a simple analytical thermal model for simple solenoids that allows one to estimate beforehand how well the coil will follow the prescribed profile. A diagram of the model is shown below. Each layer is represented by an RC element where R is the thermal resistance between layers and C is the heat capacity of a layer.

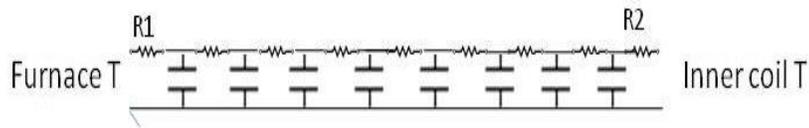

Coil Thermal Model:

1. Each RC combination is a layer in a long solenoid.
2. R is the heat conductivity in j /cm² /°C due to coil insulation.
3. C is the heat capacity of a unit area of a given layer, mainly conductor.
4. R1 and R2 represent the conduction between the outer and inner surfaces of the coil. Dominated by radiation

Fig 5.1. Schematic of the coil thermal model.

## 5.1 Comments on R and C.

Some comments are worth making about the model. First, note that even though the radius is changing, the RC elements are fixed. This is because the only important parameter in this model is the time constant of a layer. Since R is proportional the inverse of the area and C is proportional to the area, RC is invarient. It is also interesting to note that changing the slab model in Sec.2 into a cylinder of finite thickness and solving the problem in cylinderical coordinates gives results that are very insensitive to radius. In any case our model is at best semi-quantative and so we will ignore any radial dependence. Also we assume a very long solenoid. This is probably not good, as we saw in the OST solenoid, which is relatively short, that the coil response was shorter than predicted by the $n^2$ relationship which could be due to heat being absorbed thru the end faces.

R is determined by the insulation between layers. At $880^O$ C radiative transport thru the insulation is almost certainly dominant. This is also true for the connection to the furnace, R1 and R2. Let's make some crude estimates: At $880^O$ C the differential radiative transfer rate can be obtained by differentiating the Stefan-Boltzman equation which gives 0.026 watts /cm$^2$ / $^OK$ at At $880^O$ C for two black surfaces and is proportional to $T^3$. A 10 layer coil with 1 mm wire has about 0.5 grms / cm$^2$ of silver and with a Cp of about 0.3 j/grm, the heat necessary to raise a 1 cm$^2$ layer of coil 1 $^OK$ is about 0.15 j. The ramp rate thru the melting stage is 50 $^OK$/hour or .014 $^OK$/sec and therefore for ten layers a heat input of 10 x 0.15 x .014 = 0.021 watts / cm$^2$ is required which is easily supplied by radiation with a coil surface to furnace temperature difference of the order of a degree. Inside the winding, things are made much more complicated by the fiber insulation where there is competition between percolation thru the structure by radiation and conduction thru the fiber maze.

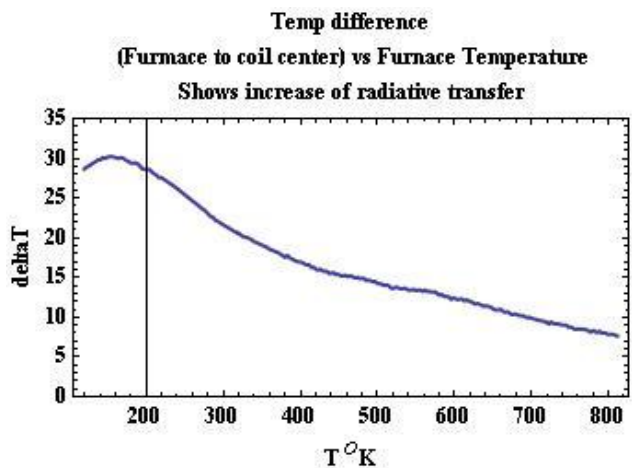

Fig. 5.2. The plot shows the temperature difference between the coil and the furnace during the initial heating ramp.

The above figure, 5.2, shows the difference between the oven temperature and the center of the NHMFL coil during the initial temperature ramp vs T °C and the decrease in the temperature difference due to radiation is evident.

Additional evidence that radiation is playing a major role is furnished by examining the difference in temperature between the coil axis and the outside temperature. The inner region of the coil is self shielded and reflects not only the furnace temperature but also that of the coil inner surface. The two figures below show this difference for the OST and NHMFL coils. If the inner temperature were anchored to the furnace, the difference would be zero. Instead both curves show clearly the melting and freezing of the 2212, with the 10 layer coil showing a much smaller effect.

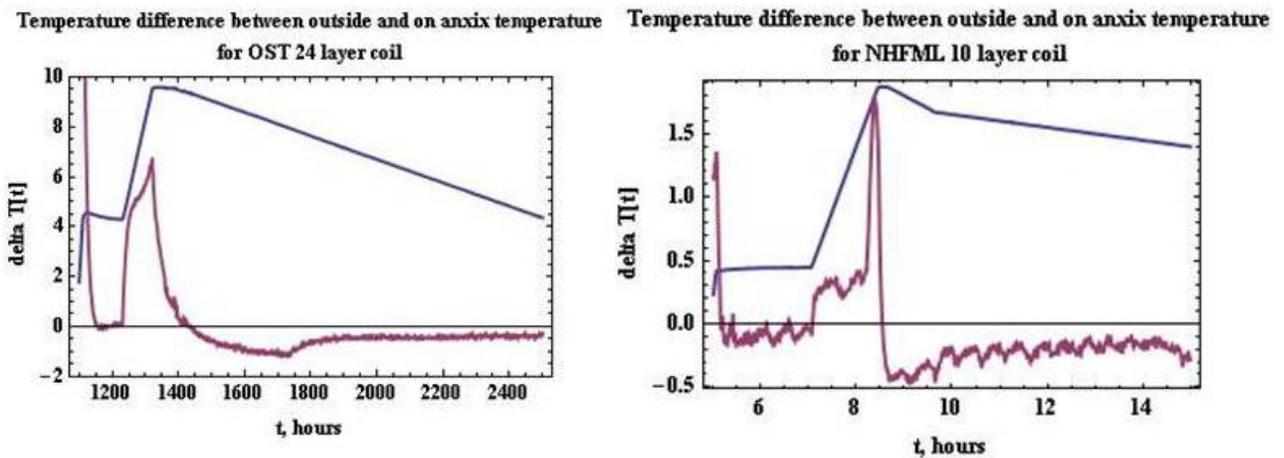

Fig. 5.3. Comparison of the difference between the outside and on axis temperature for the 24 layer and 10 layer coils. The temperature profile is not to scale and is for reference only.

The temperature of the inner surface of the OST coil is apparently better shielded from the furnace by the coil itself and hence the inside shows a large ramp difference. The NHMFL coil seems to have much better coupling to the furnace and shows a much smaller ramp difference. Both coils show the effect caused by the melting and freezing of the 2212.

The heat capacity C in the model is known over the region up to the melting point of the BSCCO. At 880 °C the Cp of silver is about 0.32 j/grm. However, when the 2212 melts, we need to add in a delta function whose area is equal to the heat of fusion times the mass of the BSCCO. This follows the approach taken with the toy model in Sec.2. Unfortunately, there is some question about the heat of fusion of the 2212. We will first of all fit the data while keeping the heat of fusion as a variable and then compare the value from the fit with what information is available.

## 5.2 The Mathematical model.

We can now write down the differential equations for the schematic shown in Fig.5.1. We assume that C[T] can be approximated by a constant plus a Gaussian located at the MP of 2212 and an area that corresponds to the heat of fusion. In the equations below, $T_{F1}$ and $T_{F2}$ give the outside and inside furnace temperatures as a known function of time. There are N layers and the temperature at t=0 must be given as an initial value for each layer. The simplest case is to start from a plateau where they are all equal.

```
C[T₁] ∂ₜT₁ =( T_F1- T₁)/R₁  -  ( T₁- T₂)/R ,
C[T_N] ∂ₜT_N =( T_N-1- T_N)/R  -  ( T_N- T_F2)/R₂ ,
C[T_n] ∂ₜT_n =( T_n-1- T_n)/R  -  ( T_n- T_n+1)/R ,
Initial values of T_n at t=0
```

Now we note that if we multiply thru by R, we get the time constant $\tau$ = RC as a parameter on the left hand side. Thus the solution is specified by the furnace temperatures, the time constant $\tau$=RC[T] and heat of fusion. Below the MP of the 2212, $\tau$ is constant determined by Cp of the wire and the resistance of the insulation. A plot of RC[T] that fits the NHMFL 10 turn coil well is shown below.

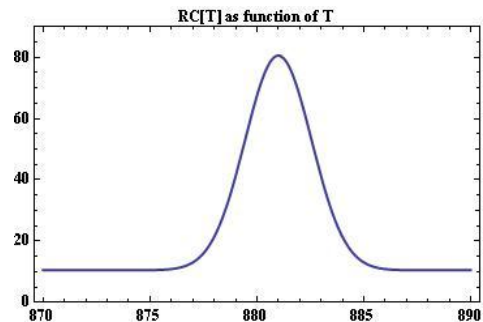

Fig. 5.4 The value of $\tau$=RC[T] that makes a good fit to the NHMFL 10 turn coil

The value of the heat of fusion will determine the length of the step seen at the MP of the 2212. The base value of $\tau$ shown in Fig. 5.4 is 11 seconds and the amplitude and width of the Gaussian were determined by fitting the above equations which easily solved numerically in Mathematica to the NHMFL data.. Note that $\tau$ = RC is not the overall time constant of the coil that we have discussed for the coils in the previous analysis. Rather it is the time constant of a single layer and is compounded into a much larger number in a multi layer coil. The furnace functions can be piece wise linear functions or can be inserted numerically form the actual measured furnace TCs. Mathematica allows the integration to be done numerically and one can start the integration any place in the cycle by giving a start time and initial value temperatures. Even if the initial conditions are wrong, the solution will relax to the correct value after a few coil time constants.

The next two figures, 5.5 and 5.6, show how well the model can fit the data over various regions of the temperature profile. The constants used are given in the table:

| Coil | Base $\tau$ | Gaussian Peak $\tau$ | Gaussian sigma | Coil Response Time |
|---|---|---|---|---|
| NHMFL   10 layers | 10.5 seconds | 70 seconds | 1.55 $^\circ$C | 2.88 minutes |
| OST         24 Layers | 6.0 seconds | 25 Seconds | 1.77 $^\circ$C | 12 minutes |

The OST coil has a response time that is not equal to the turns ratio squared times the NMFML coil partly because the layer response is faster. This could be either that the insulation between layers was different or that the aspect ratio of the two coils was different which allowed additional heat penetration from the end surfaces for the OST coil.

There are two aspects of the fit below the melting point. The first is the response time of the different layers as measured by the amount a layer lags behind the furnace ramp. The second check is how well this time constant fits the exponential match to the transition between a ramp and a plateau. The top left hand plots in figures 5.5 and 5.6 show the fit around a transition and exit from the plateau at around 825$^\circ$C. The right hand plots shows the time response at the end of the whole cycle when the furnace is ramped down.

The two bottom plots in each figure show the behavior around the melting point of 2212. The OST coil shows the melting in a more dramatic fashion primarily because the $T_{max}$ of over 890 is well above the MP of 2212. The OST coil has a lower maximum temperature and a much slower time constant. As a result complete melting does not take place till near the end of the $T_{max}$ plateau. The plots at the bottom left each show the difference between an inner TC and the furnace and the plots on the lower right show the actual temperatures. The ramp region and $T_{max}$ regions are delineated by the vertical lines.

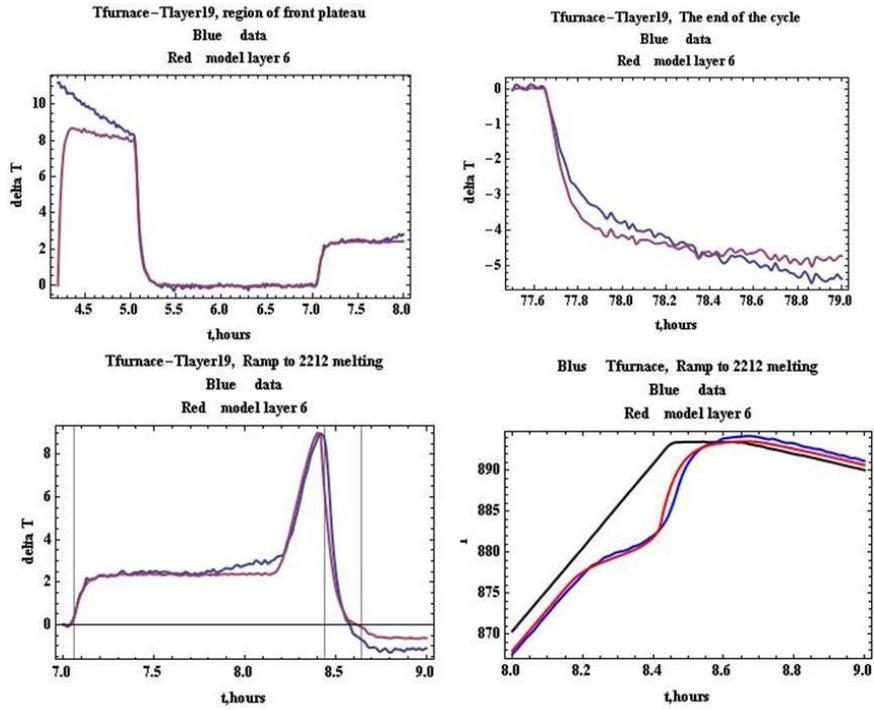

Fig. 5.5. NHMFL 10 layer coil. Data is blue and model fit is red.

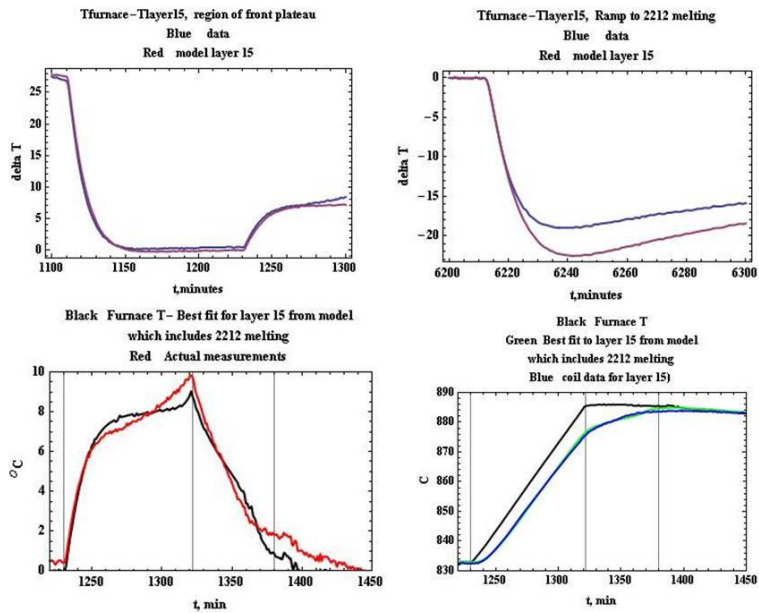

Fig 5.6 OST 24 layer coil.

## 5.3 The Heat of Fusion of 2212.

The key property controlling the region around the melting point of the 2212 is its heat of fusion. Reference [4] has a measurement but the value is unrealistically large (of the order of 1000 joules / grm) and when inserted into the model gave unrealistic results. It is actually possible to calculate the heat of fusion from the measured coil data. This is not accurate and there are some features of the fit that are not understood. The basis of the calculation is very similar to that employed in DTA measurements of the thermal properties of a material. On a linear ramp, where a piece of the strand is increasing its temperature at a steady rate, we can calculate the watts of heat influx since we know the heat capacity of the strand and $dT/dt$. (Note: to do this we also need the $C_p$ of the 2212, the amount of 2212 and the amount of Ag all at high temperature.. Unfortunately, I also haven't been able to find a value for $C_p$ of 2212 and have set it equal to 1.5 times that of silver for use here. Since the amount of 2212 is only about 14% of the total, this probably isn't a major error, but needs to be fixed!) Since we fit $RC[T]$ over the whole region and since R is a constant, this is equivalent to knowing the shape of $C[T]$ thru the whole region and we can normalize to the $C_p$ of the strand composition well away from the melting point. Then area under the Gaussian part of the curve directly gives the heat of fusion of the fractional amount of 2212 in the sample. For 1 mm strand, I have used 14.6% by total weight for the amount of 2212. The measurements shown above yield the numbers shown below:

| Heat of fusion | |
|---|---|
| Best measurement | 46 joules/grm |
| OST coil | 40 joules/grm |
| NHMFL coil | 56 joules/grm |

The value of 46 j/grm in the table above has been supplied generously supplied by Mark Rikel (3) of Nexans. The fit is better than it looks since it involves fitting the model curves to the experimental data, and the exact amount of 2212 is not known nor is its exact $C_p$ at high temperatures.

## 6.0 Conclusions

1. A simple model has been developed that should be useful for predicting the dynamic thermal behavior of coils as they become bigger.
2. The thermal time of response of a coil is easily measured by the time delay of the inside temperature of the coil relative that of the furnace during a uniform ramp.
3. The melting point of the 2212 is clearly seen and gives a self calibration of the TCs and thus it is not necessary to rely on their absolute calibration. Wrapping a short piece of conductor strand around a TC allows an independent observation of the melting point in situ and has been tried successfully at two places (FNAL, Shen and LBNL, Godeke. Private Communication). A differential arrangement of two TC as is done in DTA might also be explored. It would give a peak at the melting temperature.
4. There is a small mystery near the melting point. The curves deviate before the 881 $^O$C. Perhaps the Cp of 2212 increases rapidly before that temperature. See Fig. 5.5, lower left.
5. The freezing point of the 2212 can be seen and could be used to actually control the heat treatment profile if this were shown to be important for achieving a high Ic. For instance running a current thru the winding and using $I^2$ R heating to finely modulate the passage thru freeze out might be an interesting procedure to explore.
6. A mathematical model is presented that allows rapid analysis of coil configurations. The program includes the effect of the heat of fusion of the 2212 and we seem to require a heat of fusion that agrees with the new measurements. The Mathematica program that does this is available.

## Acknowledgements

The author received important help from Mohammad Alshar'o of Muons, Inc. in developing an ANSYS program that could handle the most general shape for a coil. The question of heat treatment of large 2212 coils originally arose from the requirement of 30 – 40 T solenoids in the final stages of a muon collider cooling channel. The experimental work of Ulf Torciewitz at NHMFL within the VHFSMC collaboration inspired the inclusion of the heat of fusion into the model. Yibing Huang very generously supplied data on the 24 layer coil from OST. Finally Eric Hellstrom from NHMFL and Mark Rikel from Nexans unraveled the heat of fusion mystery. The VHFSMC Collaroration provided the environment that made the work possible.

.

ACKNOWLEDGEMENT

Operated by Fermi Research Alliance, LLC under Contract No. DE-AC02-07CH11359 with the United States Department of Energy.